\documentclass[12pt]{article}
\pdfoutput =1
\usepackage{float} 
\usepackage{amsmath}
\usepackage{amsfonts}   
\usepackage{amssymb}

\begin{document}
\date{}
\title{{\bf{\Large Holographic droplets in p-wave insulator/superconductor transition}}}
\author{
 {\bf {\normalsize Dibakar Roychowdhury}$
$\thanks{E-mail: dibakar@bose.res.in, dibakarphys@gmail.com}}\\
 {\normalsize S.~N.~Bose National Centre for Basic Sciences,}
\\{\normalsize JD Block, Sector III, Salt Lake, Kolkata-700098, India}
\\[0.3cm]
}

\maketitle
\begin{abstract}
 In the present paper, based on the notion of marginally stable modes of vector perturbations in the AdS soliton back ground we investigate the magnetic field effect on holographic $ p $- wave insulator/superconductor transition in the probe limit. We perform explicit analytic calculations considering both Schwarzschild AdS soliton and Gauss Bonnet AdS soliton back grounds and obtain a unique relation between the chemical potential and the magnetic field near the critical point. We also extend our analysis for $ p+ip $- wave back ground in the presence of external magnetic field. In both the cases it is observed that the non abelian model exhibit \textit{droplet} solutions. Moreover it is also found that the increase in the value of the external magnetic field essentially reduces the size of the droplet and thereby makes the condensation harder.
\end{abstract}

\section{Introduction}
 
 For the past few years AdS/CFT duality \cite{ref1}-\cite{ref3} has received significant attention due to its several remarkable applications  in various areas of physics particularly towards the understanding of several strongly coupled phenomena in the usual condensed matter systems. Among its various remarkable achievements, the gauge/gravity duality has successfully explained various features of so called high $ T_c $ superconductivity in the usual condensed matter physics. These are known as holographic superconductors \cite{ref4}-\cite{ref6}. The main idea behind such theories lies in the fact that if the temperature of the black hole in the AdS space falls below certain critical temperature ($ T<T_c $) the black hole would become unstable to develop a nontrivial hair profile through the mechanism of abelian symmetry breaking in the bulk gravity theory. This essentially triggers superconductivity/super-fluidity in the boundary CFT through the mechanism of spontaneous breaking of a global (or sometimes weakly gauged) $ U(1) $ symmetry. For $ T>T_c $ the black hole possess no hair and the corresponding gravity theory stands for the dual of a conductor. Therefore these models essentially describe conductor/superconductor transition in the boundary CFT \cite{ref7}-\cite{newref3}. 
 
 Besides having the usual description of conductor/superconductor phase transition, one may also generate so called insulator/superconductor phase transition in the boundary CFT through the appropriate modifications of the bulk gravity theory, i.e; considering the gravitational theory in an AdS soliton back ground \cite{ref42}. The AdS soliton that has been first investigated in \cite{ref43} could be obtained through a double wick rotation of the usual AdS Schwarzschild solution. In the language of gauge/gravity duality the AdS soliton is dual to a confined field theory with a mass gap that essentially corresponds to an insulating phase \cite{ref44}. In \cite{ref42}, for the first time, it had been argued that in the presence of a chemical potential ($ \mu $) in the AdS soliton back ground, apart from having the usual first order insulator/conductor phase transition between AdS soliton and AdS black hole \cite{ref45}, one may have an additional (second order) phase structure through the appropriate tuning of $ \mu $. It has been observed that for $ \mu>\mu_c $, where $ \mu_c $ is some critical value of the chemical potential, the AdS soliton back ground becomes unstable to develop a non trivial profile of hair which could be interpreted as insulator/superconductor phase transition in the language of boundary field theory. This is kind of quantum phase transition where the chemical potential ($ \mu $) plays the role of doping ($ x $) in the ordinary material, like cu-prates.

 Since the discovery of such an exciting phase structure, till date a number of investigations have been performed in various directions. In \cite{ref46},\cite{ref47} the analysis was further extended taking into account the back reaction of the matter fields. It was observed that one may generate a completely new type of phase structure depending on the strength of the back reaction.  In order to back up the previous numerical results \cite{ref48}, full analytic calculations were carried out in \cite{ref49},\cite{ref50} both for the Scwarzschild AdS and Gauss-Bonnet AdS back grounds. In \cite{ref51} the authors studied the insulator/superconductor phase transition via Stuckelberg mechanism and unveiled reach physics in various phase transitions. Response to a Wilson line on the circle and to a magnetic field perpendicular to the non compact directions have been explored in \cite{ref41} where the authors discover Aharonov-Bohm like effects during insulator/superconductor transition. A holographic model of SIS Josephson junction have been successfully constructed in \cite{ref52} and the dependence of the maximal current on the width of the junction have been investigated. Based on the notion of marginally stable modes of scalar/vector perturbations in the AdS Schwarzschild background, the authors in \cite{ref53},\cite{ref54} find that these modes can actually indicate the onset of insulator/superconductor transition which is also compatible with earlier observations made by Gubser \cite{ref9},\cite{ref21}. Apart from these studies, very recently some research papers were written on insulator/superconductor transition in the context of non linear electrodynamics as well as entanglement entropy\cite{ref55}-\cite{ref56}.
 
From the above list of references, it is quite interesting to note that most of the papers that are written so far are mostly based on holographic $ s $- wave model. On the other hand, a surprisingly lesser effort is given towards the study of non abelian model of insulator/superconductor transition \cite{ref57}, particularly the magnetic response in non abelian model of insulator/superconductor transition is still unexplored. The present paper aims to fill up this gap along with addressing some other major issues from a different perspective. Based on \textit{analytic} calculations, the present work reveals the importance of marginally stable modes of vector perturbations (both for the Schwarzschild AdS and Gauss-Bonnet AdS soliton back grounds) in order to construct non abelian superconducting droplets (solutions that are confined with in a finite region and decays rapidly at large distances) during insulator/superconductor transition in presence of an external magnetic field. It is noted that the size of the droplet gradually reduces to zero as the strength of the magnetic field is increased. Furthermore, it is found that near the critical point of phase transition there exists a simple algebraic relation between the chemical potential ($ \mu $) and the magnetic field ($ B $) which indeed reveals the fact that the increase in the strength of the magnetic field makes the condensation harder to form. On top of it, for Gauss-Bonnet gravity we find this simple relation to be dependent on the Gauss-Bonnet coupling ($ \alpha $) which is also an interesting observation in itself.

It interesting to note that the droplet solutions that are obtained throughout this paper is in line with the earlier observations during conductor/superconductor transitions where both the vortex and droplet solutions have been constructed for the $ s $- wave model \cite{ref34},\cite{ref36}-\cite{ref37}. From the technical point of view the absence of vortex states in the non abelian model could be explained as follows: In the non abelian holographic model the Maxwell field enters as a $ U(1) $ subgroup of the full $ SU(2) $ symmetry group. Therefore it is not just simply coupled with the condensed charged fields through gauge covariant couplings like in the $ s $ or $ d $- wave case. Thus technically speaking the nature of this difference lies in the nature of the minimal coupling. For real life materials the high $ T_c $ type II superconductivity is believed to be generated through a $ d $- wave pairing mechanism which results into vortex states when magnetic field is applied externally. There are also real world $ p+ip $-wave models that exhibit vortex solutions but that are clearly different from the holographic $ p+ip $- wave model studied in this paper. On the other hand, one could interpret our model as a holographic model of super-fluidity and try to find connections with real life super-fluid droplets \cite{ref58}.

 Before going further, let us briefly mention about the organization of the paper. In section 2, we discuss the importance of marginally stable modes of vector perturbations in order to study non abelian superconducting droplets. The calculations are carried out both for the Schwarzschild AdS and Gauss Bonnet AdS solitonic back grounds. In section 3, the droplet solutions have been constructed for the $ p+ip $- wave back ground. Finally, the paper is concluded in section 4.

 \section{Droplet solution via marginally stable modes}
 
 By marginally stable modes in a (black hole) space time we mean perturbations in space time that do not back react on other fields and depend only radial coordinates ($ r $). In other words, marginally stable modes are essentially the quasi normal modes whose frequencies go to zero ($ \omega = 0 $) near the critical point of the phase transition. The importance of marginally stable perturbations in case of holographic model of conductor/superconductor transition has been first discussed by S. S. Gubser in his pioneering papers \cite{ref9},\cite{ref21}, where he discussed both the abelian and the non abelian model of superconductors. Later on in \cite{ref54} the authors investigate the role of marginally stable modes in AdS soliton back grounds and find that these modes indeed indicate the onset of so called insulator/superconductor transition near the critical point.

 Motivated from the above studies in the subsequent analysis we first investigate the role of marginally stable modes of vector perturbations for the $ p $- wave back ground (where we consider perturbations only in one spatial direction $ \tau^{1}dx $) in presence of an external magnetic field and in the next section we generalize our calculations considering perturbations both in the $ \tau^{1}dx $ and $ \tau^{2}dy $ directions.

 \subsection{Schwarzschild AdS soliton} 
 
 In this section we aim to construct holographic droplets for the non abelian case based on the idea of marginally stable modes. In order to do that we first note that the five dimensional $ SU(2) $ Einstein-Yang-Mills action in presence of a negative cosmological constant may be written as \cite{ref57},
 \begin{equation}
 S = \int d^{5}x \sqrt{-g}\left[\frac{1}{2} (R -\Lambda) -\frac{1}{4} F^{a}_{\mu\nu} F^{a\mu\nu} \right] 
 \end{equation}
where $ F_{\mu\nu} $ is the field strength of the non abelian gauge field. 

Before proceed further let us first note that through out this paper calculations are performed in the probe limit and we treat marginally stable perturbations as a probe into the neutral AdS soliton which may be expressed as\footnote{We are working in the polar coordinates ($ \rho, \theta $).} \cite{ref43},
\begin{equation}
ds^{2} = \frac{dr^{2}}{r^{2}g(r)}+r^{2}(-dt^{2}+d\rho^{2}+\rho^{2}d\theta^{2})+ r^{2}g(r)d\chi^{2}
\end{equation}
 where, $ g(r) = 1-\frac{1}{r^{4}} $ and $ L=1 $.
 
 This solution can be obtained from a five dimensional AdS black hole solution followed by two successive wick rotations. The boundary CFT that we are studying actually lies in a space which has the topology $ R^{1,2}\times S^{1} $.
 
 In order to proceed further we choose the following ansatz for the gauge sector close to the critical point of the phase transition line (i,e; $ \mu\sim \mu_c $ and $ \psi\sim 0 $),
 \begin{eqnarray}
 A = \tau^{3}(\mu_c dt + \frac{1}{2} B \rho^{2}d\theta) + \psi(t,r,\chi,\rho)\tau^{1}d\rho\label{e11}
 \end{eqnarray}
where $ \tau^{3} $ could be identified as the generator of the electromagnetic $ U(1) $ subgroup of the full $ SU(2) $ gauge group. Here $ B $ is a constant magnetic field which is related to the electromagnetic vector potential. 
 
With the above choice, the equation of motion for $ \psi $ turns out to be,

\begin{equation}
\partial_r (r^{3}g(r)\partial_r \psi)-\frac{1}{r}\partial_t^{2}\psi + \frac{1}{rg(r)}\partial_\chi^{2}\psi + \frac{\mu_c^{2}}{r}\psi-\frac{B^{2}\rho^{2}}{4r}\psi = 0 \label{e1}.
\end{equation}
 
 In order to solve the above equation (\ref{e1}), we choose the ansatz of the following form,
 \begin{equation}
 \psi = F(r,t)H(\chi)U(\rho) = F(r,t)H(\chi)\frac{2}{\sqrt{B}}\label{e2}
  \end{equation}
where we have actually set,
\begin{equation}
U(\rho)= \rho = \frac{2}{\sqrt{B}}.
\end{equation} 
 
From the above analysis it is indeed clear that the solutions we are looking for is actually confined with in a finite circular region whose radius $ \rho\propto\frac{1}{\sqrt{B}} $. Thus for sufficiently large magnetic field the model essentially describes a superconducting droplet for which the condensate is non zero only at the origin where the core of the droplet is located \cite{ref34},\cite{ref36}-\cite{ref37}. 
 
 Substituting the above ansatz (\ref{e2}) into (\ref{e1}) we find the following set of equations,
 
 \begin{equation}
 \frac{rg(r)}{F(r,t)}\partial_r (r^{3}g(r)\partial_r F(r,t)) - \frac{g(r)}{F(r,t)}\partial_t^{2}F(r,t)+g(r)(\mu_c^{2}-B) = \kappa^{2}\label{e3}
 \end{equation}
 
 and,
 
 \begin{equation}
 \partial_\chi^{2}H = -\kappa^{2}H\label{e4}.
 \end{equation}

The solution of (\ref{e4}) may be noted as,
\begin{equation}
H (\chi) = exp(i\kappa \chi)
\end{equation}
where, we identify\footnote{We have actually an infinite tower of marginally stable modes corresponding to different values of $ n $. It is generally expected that the lowest mode of excitation ($ n=0 $) will be the first to condense and all other modes do not matter to the phase transition.} $ \kappa = 2n $ from the periodicity property\footnote{$  H(\chi) = H(\chi+\pi)$} of $ H(\chi) $. 
 
 Finally, equation (\ref{e3}) turns out to be,
 \begin{equation}
 \partial_r^{2}F(r,t) + \left(\frac{3}{r}+\frac{g^{'}(r)}{g(r)} \right)\partial_r F(r,t)-\frac{1}{r^{4}g(r)}\partial_t^{2}F(r,t)+\frac{1}{r^{4}g}\left( \mu_c^{2}-B-\frac{4n^{2}}{g(r)}\right)F(r,t) = 0\label{e5}.  
 \end{equation}
 
 As a next step, we substitute $ F(r,t) = exp(-iwt)R(r) $, which essentially reduces (\ref{e5}) into a radial equation in the variable $ R(r) $,
 \begin{equation}
 \partial_r^{2}R(r)+\left(\frac{3}{r}+\frac{g^{'}(r)}{g(r)} \right)\partial_r R(r)+\left(\omega^{2}+\mu_c^{2}-B-\frac{4n^{2}}{g(r)} \right) \frac{R(r)}{r^{4}g(r)} = 0\label{e6}.
\end{equation}  
 
 Since in the present analysis we are interested in perturbations that are marginally stable, therefore we set $ \omega = 0 $. Also we redefine our variable as $ z=1/r $. With this redefinition equation (\ref{e6}) turns out to be,
 
\begin{equation}
\partial_z^{2}R(z)+\left(\frac{g^{'}(z)}{g(z)}-\frac{1}{z} \right)\partial_z R(z)+\left( \mu_c^{2}-B-\frac{4n^{2}}{g(z)}\right)\frac{R(z)}{g(z)} = 0 \label{e7}.
\end{equation}

 As a next step, we introduce a trial function $ \Omega (z) $ as,
 \begin{equation}
 R(z)|_{z\rightarrow 0} \sim \langle \mathcal{O}_2\rangle z^{2} \Omega (z)\label{e8}
 \end{equation}
 with the boundary conditions for $ \Omega (z) $ as $ \Omega (0) = 1 $ and $ \Omega^{'}(0)=0 $.
 
 Substituting (\ref{e8}) into (\ref{e7}) the equation of motion for $ \Omega (z) $ turns out to be
 \begin{equation}
 \partial_z^{2}\Omega(z) + p(z)\partial_z \Omega(z) + q(z) \Omega(z) + \lambda^{2}w(z)\Omega(z) = 0\label{e9}
 \end{equation}
 where we have set $ \lambda^{2} = \mu_c^{2}-B $ along with,
 \begin{eqnarray}
 p(z)&=&\frac{3-7z^{4}}{z(1-z^{4})}\nonumber\\
 q(z)&=&\frac{-8z^{2}}{1-z^{4}}-\frac{4n^{2}}{g^{2}(z)}\nonumber\\
 w(z)&=& \frac{1}{1-z^{4}}.
 \end{eqnarray}
 
 It is now easy to convert (\ref{e9}) into the standard Strum-Liouville (SL) eigen value equation \cite{ref14},
 \begin{equation}
 \partial_z(T(z)\Omega^{'}(z)) - Q(z)\Omega(z) + \lambda^{2}P(z)\Omega(z) = 0
 \end{equation}
 with the following identifications,
 \begin{eqnarray}
 T(z)&=&z^{3}(1-z^{4})\nonumber\\
 Q(z)&=&\frac{z^{3}(8z^{2}(1-z^{4})+4n^{2})}{1-z^{4}}\nonumber\\
 P(z)&=&z^{3}.
 \end{eqnarray}
 
 The corresponding eigenvalue ($\lambda^{2}$) is obtained through minimizing the following functional, 
 
 \begin{equation}
 \lambda^{2}[\Omega(z)] = \mu_c^{2}-B =\frac{\int_{0}^{1}dz(T(z)(\Omega^{'}(z))^{2}+Q(z)\Omega^{2}(z))}{\int_{0}^{1}dz P(z) \Omega^{2}(z)}\label{e10}
 \end{equation}
 
 In order to estimate the R.H.S. of (\ref{e10}) we take $ \Omega(z) = 1- a z^{2} $, which finally yields,
 \begin{equation}
 \mu_c^{2}-B = 5.13933
 \end{equation}
 with $ a = 0.338114 $.
 
 This is an important result which states that the chemical potential ($ \mu\sim \mu_c $) and the magnetic field ($ B\sim B_c $) may be connected to each other via a simple algebraic relation near the critical point of the $ p $- wave insulator/superconductor phase transition. Because of the presence of the negative sign it is easy to note that the increase in the value of the external magnetic field essentially increases $ \mu_c $ which results in a harder condensation. 
 
 In the next section we wish to verify this relation for the Gauss-Bonnet AdS soliton back ground and we will try to examine the effect of so called higher derivative coupling on the non abelian condensate in presence of external magnetic field.


\subsection{Gauss-Bonnet AdS soliton}

It has been noted earlier in \cite{ref50}, that the Gauss-Bonnet coupling ($ \alpha $) indeed affects the insulator/superconductor transition.  Increase in the value of $ \alpha $ essentially increases $ \mu_c $ which results in a harder condensation. In the present section we aim to investigate the situation in presence of an external magnetic field. The motivation of the present calculation is to see how the simple algebraic relation that is obtained in the previous section now depends on the value of the coupling $ \alpha $.

We start with the five dimensional AdS soliton in the Gauss-Bonnet gravity which (in polar coordinates) takes the following form \cite{ref50},
 \begin{equation}
ds^{2} = \frac{dz^{2}}{z^{2}g(z)}+\frac{1}{z^{2}}(-dt^{2}+d\rho^{2}+\rho^{2}d\theta^{2})+ \frac{g(z)}{z^{2}}d\chi^{2}
\end{equation}
 where, $ g(z) = \frac{1}{2\alpha}\left[ 1-\sqrt{1-4\alpha(1-z^{4})}\right]  $ with $ \alpha $ as Gauss-Bonnet coupling.

The asymptotic form of the metric is given by,
\begin{equation}
g(z)|_{z\rightarrow 0}\sim\frac{1}{2\alpha}\left[ 1-\sqrt{1-4\alpha}\right]
\end{equation}
from which we identify the effective AdS length to be,
\begin{equation}
L_{eff}^{2} = \frac{2\alpha}{1-\sqrt{1-4\alpha}}.
\end{equation}

We take the same ansatz (\ref{e11}) for the gauge field ($ A_\mu $) and set,
\begin{equation}
\psi(t,z,\chi,\rho) = R(z)S(\chi,\rho)exp(-i\omega t).
\end{equation}

We proceed in the same way as we did in the previous section. Since we are interested in marginally stable modes therefore we will be interested in the equation of motion for $ R(z) $. Finally we set $ w=0 $, which yields,
\begin{equation}
\partial_z^{2}R(z)+\xi(z)\partial_z R(z)+\left[ \mu_c^{2}-B-\frac{4n^{2}}{g(z)}\right] \frac{R(z)}{g(z)} = 0\label{e12}
\end{equation}
where, $ \xi(z)= \frac{1-4\alpha-4\alpha z^{4} - \sqrt{1-4\alpha(1-z^{4})}}{z\left[\sqrt{1-4\alpha(1-z^{4})}-1+4\alpha(1-z^{4}) \right] } $.

Our next step would be to convert (\ref{e12}) into the standard SL eigen value equation. In order to do that we first set 
\begin{equation}
 R(z)|_{z\rightarrow 0} \sim \langle \mathcal{O}_2\rangle z^{2} \Omega(z)
 \end{equation}
which yields,
\begin{equation}
 \partial_z^{2}\Omega(z) + p(z)\partial_z \Omega(z) + q(z) \Omega(z) + \lambda^{2}w(z)\Omega(z) = 0\label{e13}
 \end{equation}
 where we note,
 \begin{eqnarray}
 p(z)&=&\frac{3\sqrt{1-4\alpha(1-z^{4})}-3+12\alpha - 20\alpha z^{4}}{z\left[\sqrt{1-4\alpha(1-z^{4})}-1+4\alpha(1-z^{4}) \right] }\nonumber\\
 q(z)&=&\frac{-16\alpha z^{2}}{z\left[\sqrt{1-4\alpha(1-z^{4})}-1+4\alpha(1-z^{4}) \right] }-\frac{4n^{2}}{g^{2}(z)}\nonumber\\
 w(z)&=& g^{-1}(z)
\end{eqnarray}
with, $  \lambda^{2}=\mu_c^{2}-B $.

This equation (\ref{e13}) may be converted into a standard SL eigen value equation,
\begin{equation}
 \partial_z(T(z)\Omega^{'}(z)) - Q(z)\Omega(z) + \lambda^{2}P(z)\Omega(z) = 0
 \end{equation}
 with the following identifications,
 \begin{eqnarray}
 T(z)&=&\frac{z^{3}}{2\alpha}\left[ 1-\sqrt{1-4\alpha(1-z^{4})}\right] = z^{3}(1-z^{4})(1+\alpha(1-z^{4})) + \mathcal{O}(\alpha^{2})\nonumber\\
 Q(z)&=&-T(z)q(z)= 8z^{5}(1+2\alpha (1-z^{4})) + \mathcal{O}(\alpha^{2})\nonumber\\
 P(z)&=& T(z)w(z)=z^{3}.
 \end{eqnarray}
Since the value of the higher derivative coupling ($ \alpha $) is very small therefore for the present analysis it will be sufficient to  consider the leading possible effect of the higher derivative correction to the onset of $ p $- wave insulator/superconductor transition.

The eigen value could be obtained by varying the following functional,
 
\begin{equation}
 \lambda^{2}[\Omega(z)] = \mu_c^{2}-B =\frac{\int_{0}^{1}dz(T(z)(\Omega^{'}(z))^{2}+Q(z)\Omega^{2}(z))}{\int_{0}^{1}dz P(z) \Omega^{2}(z)} = \zeta(\alpha, n)\label{e14}
 \end{equation}

In order to estimate the above functional integral we consider the trial function to be of the form $ \Omega(z) = 1-az^{2} $, where $ a $ is determined through the minima of the functional. In the following (Table 1) we tabulate various values of $ \zeta(\alpha,n=0) $ corresponding to different choice of the coupling parameter ($ \alpha $).

\begin{table}[htb]
\caption{Values of $ \zeta $ for different choices of $\alpha$}   
\centering                          
\begin{tabular}{c c c c}            
\hline\hline                        
$\alpha$ & $ a $&  $\zeta_{SL}\left(=\mu_c^{2}-B\right)$ &  \\ [0.05ex]
\hline
0.0001 & 0.33809 & 5.13978  \\
0.1 &0.314773 & 5.59003 \\
0.2 & 0.29277 &6.03793   \\ [0.5ex] 
\hline                              
\end{tabular}\label{E1}  
\end{table} 

From the above table (1) it is indeed clear that in the presence of higher derivative corrections to the bulk action, the simple algebraic connection between the chemical potential ($ \mu_c $) and the magnetic field ($ B $) is now dependent on $ \alpha $. Furthermore we note that given a fixed value of $ B $, the corresponding value of $ \mu_c $ increases as we increase the value of the Gauss-Bonnet coupling ($ \alpha $). This indeed suggests the onset of a harder condensation and which is expected from the earlier observations \cite{ref48},\cite{ref50}.


\section{Droplet solutions in $ p+ip $- wave back ground}

In the present section, we would like to investigate the magnetic response in holographic $ p+ip $- wave superconductors considering the probe limit.  In the new coordinate $ z=1/r $, the metric of the space time may be written as,
\begin{equation}
ds^{2}= \frac{dz^{2}}{z^{2}g(z)}+\frac{1}{z^{2}}(-dt^{2}+dx^{2}+dy^{2})+ \frac{g(z)}{z^{2}}d\chi^{2}
\end{equation}
where, the explicit form of $ g(z) $ depends on the particular choice of space time.

In this fixed back ground, the corresponding equation of motion for the gauge field turns out to be,
\begin{equation}
\frac{1}{\sqrt{-g}}\partial_\mu(\sqrt{-g}F^{a\mu\nu})+ \epsilon^{abc}A^{b}_\mu F^{c\mu\nu} = 0\label{e15}
\end{equation}
with the choice $ \epsilon^{123} = 1 $.

In order to proceed further we choose the following ansatz for the gauge field in the $ p+ip $- wave back ground as,
\begin{equation}
A=(\phi(x,y,z)dt+A^{3}_y(x,y,z) dy)\tau^{3} + w(x,y,z)(\tau^{1} dx + \tau^{2} dy)\label{e16}
\end{equation}
where the condensate of $ w $ bosons essentially breaks the original rotational invariance of the system and thereby triggering superconductivity in the boundary theory.

Substituting the ansatz (\ref{e16}) into (\ref{e15}) the following set of equations for $ \phi $, $ A^{3}_y $ and $ w $ may be respectively obtained as,
\begin{equation}
\partial_z\phi^{2}+\left(\frac{g^{'}(z)}{g(z)}-\frac{1}{z} \right)\partial_z \phi + \frac{1}{g(z)}(\partial_x^{2}+\partial_y^{2})\phi - \frac{2w^{2}\phi}{g(z)} = 0
\end{equation}
\begin{eqnarray}
\partial_z \left(\frac{g(z)}{z}\partial_z A^{3}_y \right) + \frac{1}{z}\partial_x^{2}A^{3}_{y}+\frac{3w}{z}\partial_x w-\frac{1}{z}w^{2}A^{3}_{y} = 0
\end{eqnarray}
and,
\begin{eqnarray}
\partial_z \left(\frac{g(z)}{z}\partial_z w \right)+ \frac{1}{2z}(\partial_x^{2}+\partial_y^{2})w-\frac{3w}{2z}\partial_x A^{3}_y-\frac{w^{3}}{z}+\frac{\phi^{2}w}{z}-\frac{w}{2z}(A^{3}_y)^{2} = 0\label{e17}.
\end{eqnarray} 
 
 From the structure of these equations it seems to be quite difficult to have exact analytic solutions for these equations.  However, one can always make perturbative expansion of the fields near the critical value of the magnetic field strength ($ B \sim B_c $) where there is no condensation ($ w=0 $),
 \begin{eqnarray}
 \phi &=& \phi^{(0)}+\varepsilon \phi^{(1)}+.. .. \nonumber\\
 A^{3}_y &=& A^{3(0)}_y + \varepsilon A^{3(1)}_y +.. ..\nonumber\\
 w&=&\sqrt{\varepsilon}w_1 + \varepsilon^{3/2}w_2 +.. ..\label{e18}
 \end{eqnarray}
where the perturbation expansion is carried out on the parameter $ \varepsilon = \frac{B_c - B}{B_c} $, when $ B $ is slightly below $ B_c $. Here $ \phi^{(0)} $ and $ A^{3(0)}_y $ are zeroth order solutions which we set,
\begin{eqnarray}
\phi^{(0)} =  \mu_c ~~~ and,~~~ A^{3(0)}_y = B_c x
\end{eqnarray}
in order to make the solutions consistent with the Newmann type boundary conditions for the fields \cite{ref49}.

Substituting (\ref{e18}) into (\ref{e17}), and neglecting terms\footnote{Near the critical point, $ w\sim 0 $, therefore we may ignore the quadratic terms.} $ \sim \mathcal{O}(\varepsilon w) $ we obtain,
\begin{equation}
2z \partial_z \left(\frac{g(z)}{z}\partial_z w_1 \right)+(\partial_x^{2}-B_c^{2}x^{2}-3B_c)w_1+\partial_y^{2}w_1+2\mu_c^{2}w_1 = 0\label{e19}
\end{equation}

In order to solve (\ref{e19}), we choose the following ansatz,
\begin{equation}
w_1(x,y,z) = X(x)Y(y)Z(z)\label{e20}.
\end{equation}

Finally, substituting (\ref{e20}) into (\ref{e19}), we arrive at the following set of equations,
\begin{equation}
\partial_{\tilde{y}}^{2}Y = -p^{2}\label{e21}
\end{equation}
\begin{equation}
(-\partial_{\tilde{x}}^{2}+\tilde{x}^{2}+3-k^{2})X(\tilde{x}) = 0\label{e22}
\end{equation}

and,
\begin{equation}
\partial_z^{2}Z(z)+\left(\frac{g^{'}(z)}{g(z)}-\frac{1}{z} \right)\partial_z Z(z)+\left( \mu_c^{2}-B_c\right)\frac{Z(z)}{g(z)} = 0 \label{e23}
\end{equation} 
where\footnote{We choose the constants $ k^{2} $ and $ p^{2} $ such that $ k^{2}+p^{2} = 2 $.}, we set $ \tilde{x}=\sqrt{B_c}x $ and $ \tilde{y}=\sqrt{B_c}y $.

Solutions for (\ref{e21}) and (\ref{e22}) may be expressed as,

\begin{equation}
Y\sim exp(ip\tilde{y})
\end{equation}
and,
\begin{equation}
X(\tilde{x}) = e^{-\frac{\tilde{x}^{2}}{2}}H_n(\tilde{x})\label{e24}
\end{equation}
respectively, where $ H_n(\tilde{x}) $ are standard Hermite functions.

Before we proceed further, let us pause for some time and try to extract physics from the above results. First of all, and most importantly, from (\ref{e24}) we note that the solution in the $ x $ direction does not depend on $ p $. This situation is completely different from the $ s $- wave case, where the solutions obtained for $ x $ directions essentially depend on $ p $, which results in a vortex lattice solutions with a periodicity in the $ x $ direction\cite{ref34}. Thus we see that for the non abelian insulator/superconductor transition we can not have vortex lattice solutions like we have in the abelian case \cite{ref41}. The reason for this is that in the non abelian case we have a different scenario where the the Maxwell sector appears as a $ U(1) $ subgroup of the full $ SU(2) $ symmetry group and essentially is not minimally coupled with the condensed charged fields like in the abelian case. On top of it, we see that for large values of the magnetic field, or at large distances the solution (\ref{e24}) rapidly dies out. This on the other hand suggests that these are droplet solutions. Finally, from (\ref{e23}) we note that, we may easily convert it into a standard SL form that we did earlier and as a consequence of that one can easily show $ \mu_c^{2} -B_c = constant$.


\section{Summary and final remarks}

In the present paper, based on the notion of marginally stable modes of vector perturbations in an AdS soliton back ground, several properties of non abelian model holographic insulator/superconductor transition have been investigated in presence of an external magnetic field in the probe limit. A number of interesting results have been obtained in this regard. First of all we have found that the non abelian model exhibit only droplet solutions. One of the (technical) reason for this is that in the non abelian case we do not encounter a minimal coupling between the Maxwell field and the charged condensed field like in the abelian model. As a result of this, the superposition of droplet solutions do not give rise to any vortex lattice solution. Similar situations were also encountered earlier in \cite{newref1}.

On top of it, we have found a simple mathematical relation between the chemical potential and the external magnetic field near the critical temperature. This relation in itself is very interesting since it implies that an increase in the magnetic field will always be accompanied by a corresponding increase in the (critical) chemical potential, which in fact suggests the onset of a harder condensation. Moreover, for the Gauss-Bonnet gravity we find this relation to be dependent on the Gauss-Bonnet coupling itself. In the future one can make further probe into the non abelian model of insulator/superconductor transition by studying this model in presence of back reaction and particularly one should try to see the effect of back reaction on the holographic condensate in presence of external magnetic field.

\vskip 5mm
{\bf{Acknowledgement :}}\\
     Author would like to thank the Council of Scientific and Industrial Research (C. S. I. R), Government of India, for financial help.


\end{document}